\RequirePackage{fix-cm}
\documentclass[twocolumn]{svjour3}          % twocolumn
\smartqed  % flush right qed marks, e.g. at end of proof
\usepackage{graphicx}
\usepackage{xcolor}
\journalname{myjournal}
\begin{document}

\title{Experimental investigation of the turbulent cascade development by injection of single large-scale Fourier modes\thanks{Financial support from the Poul Due Jensen Foundation (Grundfos Foundation) for this research is gratefully acknowledged. Grant number 2018-039.}
}
%\subtitle{Do you have a subtitle?\\ If so, write it here}

%\titlerunning{Short form of title}        % if too long for running head

\author{Margherita Dotti \and
        Rasmus K. Schlander \and
        Preben Buchhave \and
        Clara M. Velte
}

%\authorrunning{Short form of author list} % if too long for running head

\institute{M. Dotti \at
              Department of Mechanical Engineering, Technical University of Denmark, Nils Koppels All\'{e}, Bldg. 403, 2800 Kgs. Lyngby, Denmark. \\
              \email{mardott@kt.dtu.dk}           %  \\
             \emph{Present address:} Department of Chemical and Biochemical Engineering, Technical University of Denmark, S{\o}ltofts Plads 228A, 2800 Kgs. Lyngby, Denmark  %  if needed
           \and
           R.K. Schlander \at
              Department of Mechanical Engineering, Technical University of Denmark, Nils Koppels All\'{e}, Bldg. 403, 2800 Kgs. Lyngby, Denmark. \\
              \email{rasmuskorslund@gmail.com}           %  \\
             \emph{Present address:} Department of Aeronautics, Imperial College, London SW7 2AZ, United Kingdom %  if needed
             \and
             P. Buchhave \at
              Intarsia Optics, S{\o}nderskovvej 3, 3460 Birker{\o}d, Denmark.\\
              \email{buchhavepreben@gmail.com}           %  \\
           \and
           C.M. Velte \at
              Department of Mechanical Engineering, Technical University of Denmark, Nils Koppels All\'{e}, Bldg. 403, 2800 Kgs. Lyngby, Denmark. \\
              ORCID: 0000-0002-8657-0383 \\
              \email{cmve@dtu.dk}           %  \\
}

\date{Received: date / Accepted: date}
% The correct dates will be entered by the editor

\maketitle

\begin{abstract}
The current work presents an experimental investigation of the dynamic interactions between flow scales caused by repeated actions of the nonlinear term of the Navier-Stokes equation. Injecting a narrow band oscillation, representing a single Fourier mode, into a round jet flow allows the measurement of the downstream generation and development of higher harmonic spectral components and to measure when these components are eventually absorbed into fully developed turbulence. Furthermore, the dynamic evolution of the measured power spectra observed corresponds well to the measured cascaded delays reported by others. Closely matching spectral development and cascade delays have also been derived directly from a one-dimensional solution of the Navier-Stokes equation described in a companion paper. The results in the current work provide vital information about how initial conditions influence development of the shape of the spectrum and about the extent of the time scales in the triad interaction process, which should be of significance to turbulence modelers.
% \PACS{PACS code1 \and PACS code2 \and more}
% \subclass{MSC code1 \and MSC code2 \and more}
\end{abstract}

\section{Introduction}
\label{intro}
Most of the 20th century knowledge of turbulence is founded on the equilibrium gas dynamics analogy to small and intermediate scales of turbulence, as put forth by Kolmogorov~\cite{1}. Indeed, almost all turbulence theories and models are in one way or another based on the Kolmogorov theory of turbulence (which the authors will hereafter refer to as the K41 theory). In particular, many approaches rely on the existence of a continuous exchange of turbulent energy from small to large wave numbers, the Richardson cascade, where predominantly local interactions between scales are assumed to occur~\cite{2,3,4}.

\begin{figure}
% Use the relevant command to insert your figure file.
% For example, with the graphicx package use
  \includegraphics[width=\linewidth]{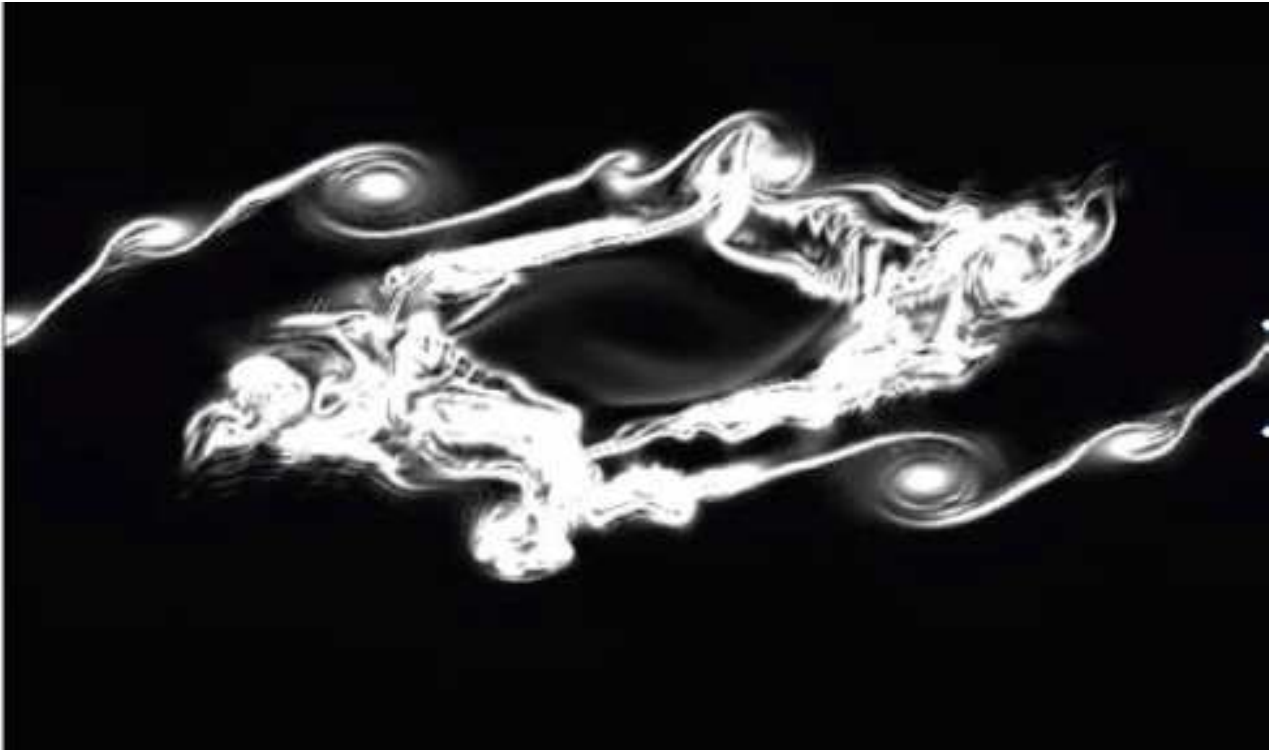}
% figure caption is below the figure
\caption{Generation and interaction of non-local scales~\cite{5}.}
\label{fig:1}       % Give a unique label
\end{figure}

By taking the Fourier transform of the non-linear advection term in the Navier-Stokes equation, one can immediately observe the possibilities of energy transfer between different wavenumbers of the velocity field. However, being able to infer the restriction of locality of the triad interactions, as postulated by Richardson, directly from the governing equations appears to have so far eluded the turbulence community. On the other hand, accumulating evidence witnesses that the actual underlying processes of energy exchange between scales may in fact be quite different from the ideas of K41. An example of non-local energy exchange, in direct violation of the Richardson cascade concept, is shown in Figure~\ref{fig:1}~\cite{5}. This figure zooms into a shear layer, from a Direct Numerical Simulation (DNS) computation, where structures of small size appear and directly exchange energy with significantly larger structures. A 2D Fourier analysis of this picture would reveal scales of vastly different size suggesting interactions through the nonlinear term in the Navier-Stokes equation.

Historically, much (but not all) of the published experimental evidence supported the K41 theory. However, George~\cite{6}, inspired by recent developments of his and others, argued that the K41 only applies to equilibrium flows, while failing to reproduce results from flows out of equilibrium. %This has now led to believe that the K41 theory constitutes an equilibrium solution that does not appear to hold true in general.

Recently, interesting dynamic effects in the transfer of turbulent energy by the triad interaction process have been reported. For instance, the data of the hot-wire anemometer analysis conducted by Josserand \textit{et al.}~\cite{7} reveals both a lack of time reversal symmetry and a delay in the triad interactions. This delay seems to depend on the separation of the k-vectors entering the process. Other relevant studies concern the comparison between the strength of non-local interactions against the local ones~\cite{8,9} and the persistence of the initial large-scale structures during the further turbulence development~\cite{10}. As Zhou~\cite{8} pointed out, and several investigations concluded, the local energy transfer resulted from non-local interactions~\cite{11,12,13,22}, which contrasts with the classical Kolmogorov picture.

Indeed, direct coupling between scales of quite different size seems to be possible, as is also shown in Figure~\ref{fig:1}. And even though the net effect of the energy transport is towards higher wavenumbers, the energy may move in the direction of both smaller and larger scales. Other outstanding questions concern the persistence~\cite{15} of the initial conditions into the developed turbulence and the precise form of the final decay of the turbulent fluctuations~\cite{16}. Moreover, the fact that many papers report deviations from the $-5/3$ slope in turbulence velocity power spectra~\cite{17,18,19} and that significant differences occur between measurements and model calculations in flows of significant stagnation, separation and transient processes (i.e. where the flows can potentially be pushed out of the assumed equilibrium) highlight the need for investigations that go beyond the Kolmogorov theory. For this reason, the authors believe that the turbulence energy transfer investigation should refer directly to the governing Navier-Stokes equation~\cite{20,21}.

Related work concerns measurement of the time of large-scale modulation transferred to a turbulent flow~\cite{20b} and ``periodically kicked'' turbulence~\cite{20c}. Also, studies of initial flow with a truncated power spectrum~\cite{20d} seems related to our work, where we inject only a single mode. The relevance to the active flow control community is apparent and injection or suction at certain frequencies is to be considered common practice (however typically for suppressing disturbance oscillations)~\cite{20e,20f,20g}.

The current work describes an attempt to isolate and visualize triad interactions and their dynamics by measuring the downstream development of a single or a few Fourier components, here performed as narrow-band temporal and spatial oscillations, injected into a well-defined flow. The downstream position is interpreted as a time development, given by the convection time for a fluid element, in which the fluid has been exposed to the actions of the Navier-Stokes equation. If Taylor's hypothesis can be invoked, the temporal and spatial development of the flow is by definition coupled. However, in general, Taylor's hypothesis cannot be assumed to be valid, in particular in high intensity flows. By including time as an independent variable in the analysis, the dynamic downstream development can be predicted~\cite{20,21}. The triad interaction mechanisms have been generated and isolated by inserting an oscillation into a mean flow direction with two different methods. The measurements have been compared with the flow development simulated by a one-dimensional solution of the Navier-Stokes equation, which is described in more details in the companion papers~\cite{20,21}. Our numerical method employs a succession of quasi-infinitesimal steps allowing us to numerically linearize the terms in the NSE while keeping its nonlinear development. The present work shows good agreement between the experiments and the analysis and computations and conclusions are drawn based on the resulting dynamics of the initial mode injection.

As the modal interactions described by the nonlinear term in the Navier-Stokes equation govern the development of turbulent flows, and since this mechanism is most clearly observed in Fourier space, we have devised experiments designed to show how a single Fourier mode (defined as a narrow-band oscillation) interacts with both a developed, high intensity turbulent flow and with a low intensity, quasi-laminar flow. In the developed case, the forcing of similar energy as the background turbulence is observed initially as a superposition with the background turbulence, that eventually gets absorbed into the background turbulence by exchange with the range of the background turbulent modes. In the nearly laminar jet core, the injected oscillations are strong in comparison to the background flow, however still following a similar behaviour as also predicted by the governing equations.

In the two companion papers that predict these nonlinear behaviours, we have described the theory for these interactions in a real-life measurement situation~\cite{20} and a computer simulation, which is able to follow the temporal development of a flow defined by an initial velocity time record~\cite{21}. The goal of the current work is to throw light on these modal interaction processes, as predicted in the companion papers, by tailored experiments.

More specifically, we expect to see the appearance of exact multiples of the injected frequency; From the repeated actions of the second order nonlinear term in the Navier-Stokes equation, the analysis in~\cite{20,21} reveals that frequency doubling and sum- and difference-frequencies should appear. These frequencies are clearly observed and seen to be quite persistent. As the doubling and sum-frequencies are, from the analysis, expected to dominate in energy over the difference-frequencies, we expect predominantly the gradual appearance of higher order harmonics as a dynamic cascade process where successively higher and higher frequencies are created. This is also apparent from our observations herein. Finally, the theory and simulations predict effects on the form of the generated power spectra due to the finite spatial and temporal measurement ranges (windows). This includes the expected suppression of odd harmonics for spatially finite flows or measurement domains, which is clearly demonstrated in the smallest jet flow measurements.

Future investigations, which are outside the scope of the current investigations, could provide further insight into outstanding challenges such as predicting and measuring the time scales of the dynamic spectral development.

\section{Navier-Stokes equation}
\label{sec:1}
\begin{figure}
% Use the relevant command to insert your figure file.
% For example, with the graphicx package use
  \includegraphics[width=\linewidth]{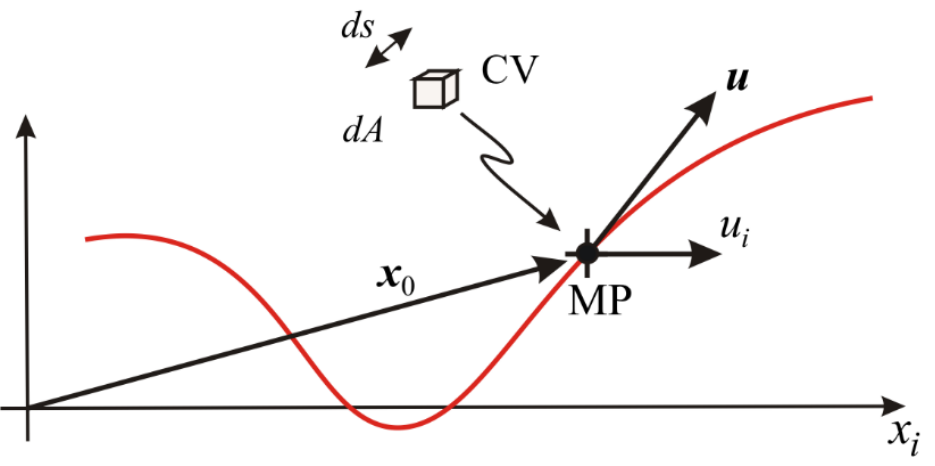}
% figure caption is below the figure
\caption{Fluid control volume and instantaneous velocity $u$ and a component $u_i$ at a point in space-time. The red line indicates a streak line.}
\label{fig:2}       % Give a unique label
\end{figure}

The Navier-Stokes equation is a three-dimensional description of the time evolution of fluid flow. The equation is essentially the application of Newton's second law to the motion of fluid passing through an infinitesimal fluid control volume, as shown in Figure~\ref{fig:2}.

The development of the flow structures and the interchange of energy between them is primarily caused by the non-linear term in the Navier-Stokes equation. This nonlinear term causes a variation in velocity, due to the change in convection of momentum through a control volume, $CV=dA\cdot ds$, where $dA$ is the cross-sectional area and $ds$ is the thickness of the $CV$. The viscous diffusion term causes a loss of velocity by diffusion of momentum to the fluid surrounding the control volume.

The pressure gradient influences the velocity by fluctuating (dynamic) pressure caused by velocity fluctuations elsewhere in the fluid and propagating instantaneously to the control volume in an incompressible fluid. This term must be computed by a full solution to the Laplace equation including pressure terms at the boundaries or inferred from theoretical models. Thus, the equation is local, describing the conditions at a `point', non-local effects only entering though pressure effects created by velocity fluctuations or through boundary conditions elsewhere in the fluid, transmitted to the point of interest.

We have designed several experiments to explore the underlying wave interaction dynamics: In the first experiment, an actual narrow-band oscillating velocity, that resembles a single Fourier mode, is injected into a fully developed turbulent jet flow. By conducting measurements downstream of the injection position, the development in time of the power spectrum is traced through laser Doppler anemometry (LDA) and hot-wire anemometry (HWA) measurements. The downstream position is interpreted in relation to the Navier-Stokes equation as a time interval in which the equations operate on the flow passing a series of control volumes, as shown in Figure~\ref{fig:2}. Tests of zither-like grid generated modes were performed as early as 1980~\cite{23}, and detailed studies of full and narrow-band velocity correlations in wind tunnel flow was made as early as 1971. However, the clear evolution of the spectrum that we report here has not been reported before.

A `full' solution to the Navier-Stokes equation requires a four-dimensional numerical solution with a high spatial and temporal resolution (DNS). However these numerical solutions do not necessarily provide a physical understanding, i.e., the comprehension of the flow based on the governing equations, which is the primary interest in the current work. Therefore, to understand the inner workings of the interactions between Fourier components (the triad interactions), a one-dimensional model has been developed and implemented~\cite{21}. The method involves projection of the forces acting on the fluid in the control volume onto the instantaneous velocity direction. The time sampling interval $\Delta t$ is then converted to the convection sampling interval $\Delta s = u \Delta t$~\cite{24}. This method, obviously, does not allow to see the full 4-dimensional motion of the fluid. However, the forces acting in the direction of the velocity will change the momentum and allow computation of kinetic energy and spatial velocity structures. The main problem is the inability of computing the pressure gradient, since this requires the solution of Laplace's equation for the whole fluid volume at the present instant in time. To include pressure, it will be necessary to invoke a model or use separate information about the fluctuating pressure.

\section{Experiments}
\label{sec:2}

The experiments described in the following are designed to inject a single Fourier mode into a well-defined flow and by measurement of the time trace along the downstream evolution follow the time development of the velocity field. As Fourier modes are essentially plane waves~\cite{24b}, the attempt was to inject a two-dimensional oscillating wave front developing along the mean flow direction.

The axisymmetric turbulent round jet can be considered ideal to investigate the turbulent cascade. Although the -5/3 power spectrum was observed already in 1962~\cite{GSM},  the round turbulent jet has proven to be an ideal flow for the study of the theory of Kolmogorov and to show the -5/3 range in the power spectrum~\cite{25}. In addition, this flow evolves rapidly enough to be practical on laboratory scales, and it is optically accessible. Moreover, it offers a useful wide literature, as it has been investigated by many over the years, see e.g.~\cite{26,27,28,29}.

The measurements are divided in three sets of experiments, presented in chronological and thought chain order, and were carried out using a laser Doppler anemometer and a hot-wire anemometer. First, LDA measurements were performed behind an oscillating airfoil designed to inject a single frequency oscillation into a high intensity turbulent round jet flow of similar strength as the injected oscillations. Secondly, measurements in jet flows of different Reynolds numbers from the same jet facility were carried out behind a larger airfoil with less vigorous oscillations using a hot-wire anemometer. Finally, the same hot-wire anemometer system was used to measure vortex shedding in a low intensity turbulent jet of significantly larger jet exit diameter, employed as an open wind tunnel. This third experiment was performed in the laminar jet core using a sharp-edged rod to create a distinct vortex shedding frequency of high intensity compared to the background flow.

\subsection{Oscillations in turbulent jet -- small airfoil (LDA)}
\label{sec:2}

\begin{figure}
% Use the relevant command to insert your figure file.
% For example, with the graphicx package use
  \includegraphics[width=\linewidth]{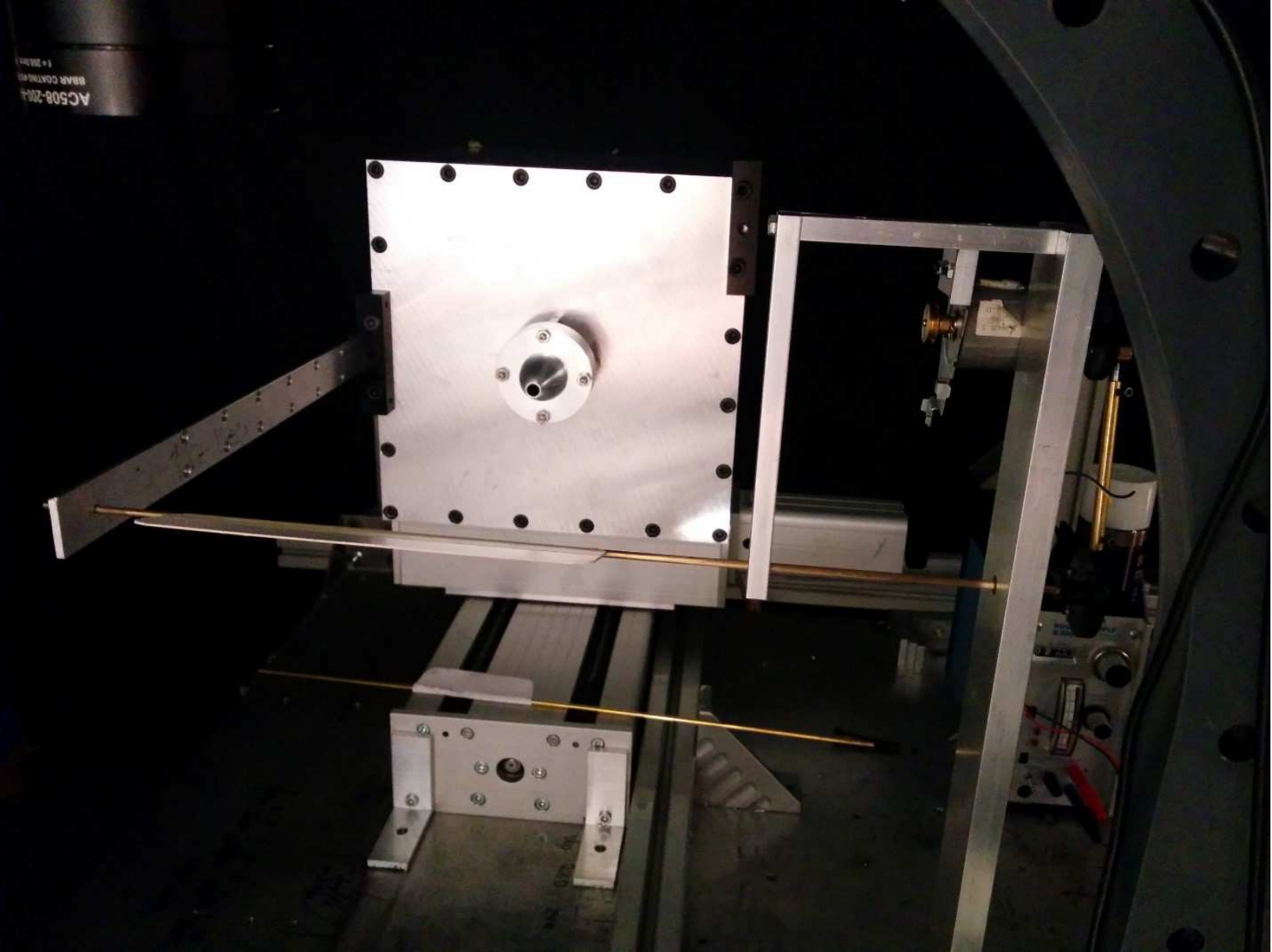}
% figure caption is below the figure
\caption{Jet generator and oscillating airfoil.}
\label{fig:3}       % Give a unique label
\end{figure}

A jet generator, which is the one used in several projects of the Turbulence Research Laboratory of DTU~\cite{30,31,32}, produces an axisymmetric fully developed turbulent flow 30 jet exit diameters downstream of the nozzle exit. The jet nozzle, shown in Figure~\ref{fig:3}, has an exit diameter of $D=10\,mm$, a contraction ratio of $3.2:1$ and can be moved in two horizontal directions, manually or by computer control. The jet is connected to a pressurized air output, used to adjust the flow velocity, and to a particle dispenser that seeds the flow. The nozzle was designed using a fifth order polynomial contraction shape as suggested by~\cite{33} in order to avoid boundary layer separation at the walls and to obtain a uniform mean flow at the outlet.

\begin{figure}
% Use the relevant command to insert your figure file.
% For example, with the graphicx package use
  \includegraphics[width=\linewidth]{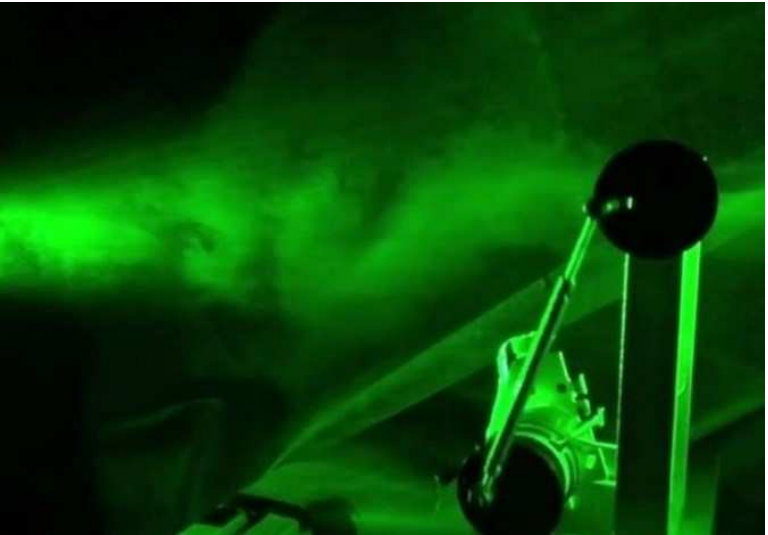}
% figure caption is below the figure
\caption{Flow visualization airfoil experiments.}
\label{fig:4}       % Give a unique label
\end{figure}

The airfoil altered the jet flow by introducing a strong sinusoidal frequency. This airfoil was situated 20 jet exit diameters downstream of the jet exit at the height of the nozzle centerline and extended through the full width of the jet, as depicted in Figure~\ref{fig:3}. Figure~\ref{fig:4} shows a smoke visualization using a vertical laser sheet across the streamwise development, illustrating the jet flow perturbed by the airfoil. Initially, the airfoil span experiment was conducted by flapping in an oscillating manner a $50 \times 10\, mm$ airfoil with a thick leading edge and a sharp trailing edge.

\begin{figure}
% Use the relevant command to insert your figure file.
% For example, with the graphicx package use
  \includegraphics{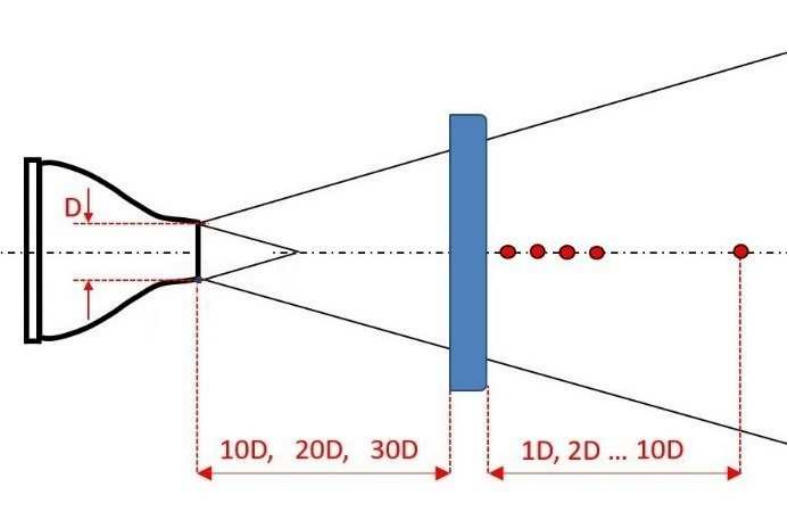}
% figure caption is below the figure
\caption{Schematic (not to scale) of the setup of the oscillating airfoil in a turbulent jet. The airfoil is depicted in blue and the measurement points in red.}
\label{fig:5}       % Give a unique label
\end{figure}

The airfoil was actuated by a small motor, creating an oscillation frequency of $10\, Hz$. The streamwise velocity component was measured using an in-house state-of-the-art side scatter LDA system~\cite{31}. Since the side scattering optics of the LDA system was highly sensitive to misalignments, it was decided to traverse the jet instead of the LDA system. The measuring distance was $300\, mm$ and the size of the measurement volume was, due to the side scattering configuration, nearly spherical with a diameter of $200\,\mu m$. When performing LDA measurements, the ambient air was seeded with glycerin particles, injected by means of pressurized air, so that a nearly uniform spatial seeding density is achieved. The size of the scattering particles ($\sim 1-5 \,\mu m$) has been shown to be small enough to faithfully track the flow and observed to be large enough to scatter enough light for a to produce a satisfactory signal quality~\cite{34}. The Doppler signal at each downstream position was measured in $400$ records, each of $2\, s$, with a sampling rate of $25\,MHz$. The Reynolds number at the jet exit was $Re_D = U_0 D / \nu = 2.2\cdot 10^4$. The LDA measurement volume was then placed at 1D, 2D, 4D, 6D, 8D, 10D jet diameters downstream from the trailing edge of the airfoil, as depicted in Figure~\ref{fig:5}.

\begin{figure*}
% Use the relevant command to insert your figure file.
% For example, with the graphicx package use
  \includegraphics[width=\textwidth]{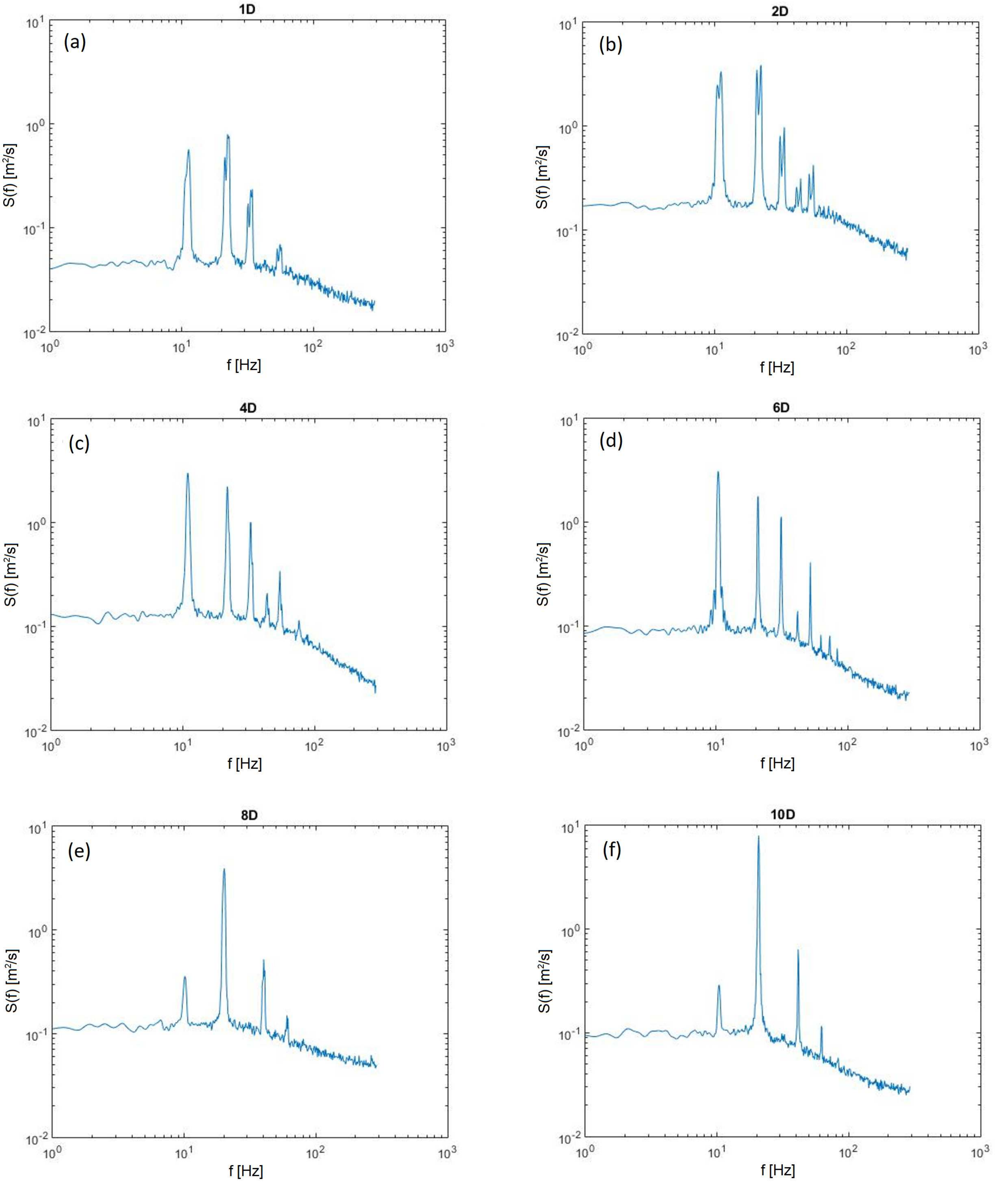}
% figure caption is below the figure
\caption{Dynamic evolution of a `wave-like behavior' of the spectral harmonic peaks for distances of $1D$, $2D$, $4D$, $6D$, $8D$ and $10D$ behind the airfoil trailing edge, respectively.}
\label{fig:6}       % Give a unique label
\end{figure*}

The corresponding measured temporal streamwise velocity power spectra are displayed in Figure~\ref{fig:6}. The sequence of plots shows interesting dynamics in the energy exchange process between scales. Energy tends to be exchanged between harmonics of the base frequency of $10\, Hz$, which is well in line with the fact that the nonlinear term in the Navier-Stokes equation introduces frequency doubling, sum- and difference-frequencies as shown analytically in~\cite{21}\footnote{This is easily observed e.g. by substituting a velocity wave traveling in the x-direction $u(x)=U \cos (k_x x)$ into $(\mathbf{u}\cdot \nabla )\mathbf{u}$}. From the figure it can also be inferred that the center of gravity of the induced spectral peaks moves towards higher frequencies as the flow is traced downstream. The alternation of the dominance of the first ($10\,Hz$) and second ($20\,Hz$) harmonic, the `wave behavior', is in agreement with Josserand \textit{et al.}~\cite{7}, who observed that energy transfers primarily among wavenumbers which are strongly correlated to the fundamental frequency. Moreover, it shows that the energy can flow in both directions, but the global energy transfer remains towards higher frequencies.

\subsection{Oscillations in turbulent jet -- large airfoil (HWA)}
\label{sec:3}

The LDA measurements and flow visualization from the previous section showed that the $50\times 10\, mm$ airfoil required a significant amplitude to create the desired modulation to ensure injection and isolation of the development of a single mode into the jet. Thus, new measurements were conducted with a larger (span and chord) airfoil, which was oscillating with a smaller amplitude to introduce less energy. The dimension of both the chord and the span dimension of the airfoil was increased to $50$ and $210\, mm$, respectively. A hot-wire anemometer could, to sufficient accuracy, be used in this less turbulent flow because the power spectrum could be processed online, and the system could be more easily traversed to cover more downstream measurement points.

A Mini CTA $54T30$ system for measurements in air from Dantec Dynamics was used for acquisition. The $55P11$ straight single wire probe (Tungsten, $d=5\, \mu m$ and length $1\, mm$) was connected through the straight $55H20$ support through a $4.0\,m$ cable. Calibration was performed across $10$ velocity points, from $1\, ms^{-1}$ to $30\, ms^{-1}$ and measuring the pressure difference with an $FCO560$ Furness Anemometer. The record length of each measurement was set to $120\, s$ and each signal was sampled with a rate of $30\, kHz$. The data was transferred to a computer passing through an anti-aliasing filter and processed with the Dantec MiniCTA $v4.05$ software. The oscillating airfoil was positioned with its leading edge at $10$, $20$ and $30$ jet exit diameters, respectively, downstream of the jet exit and different Reynolds number flows were tested. The velocity was measured at several positions downstream from the airfoil trailing edge until the point where the injected mode was fully absorbed into the background turbulence of the jet.

\begin{figure*}
  \includegraphics[width=\textwidth]{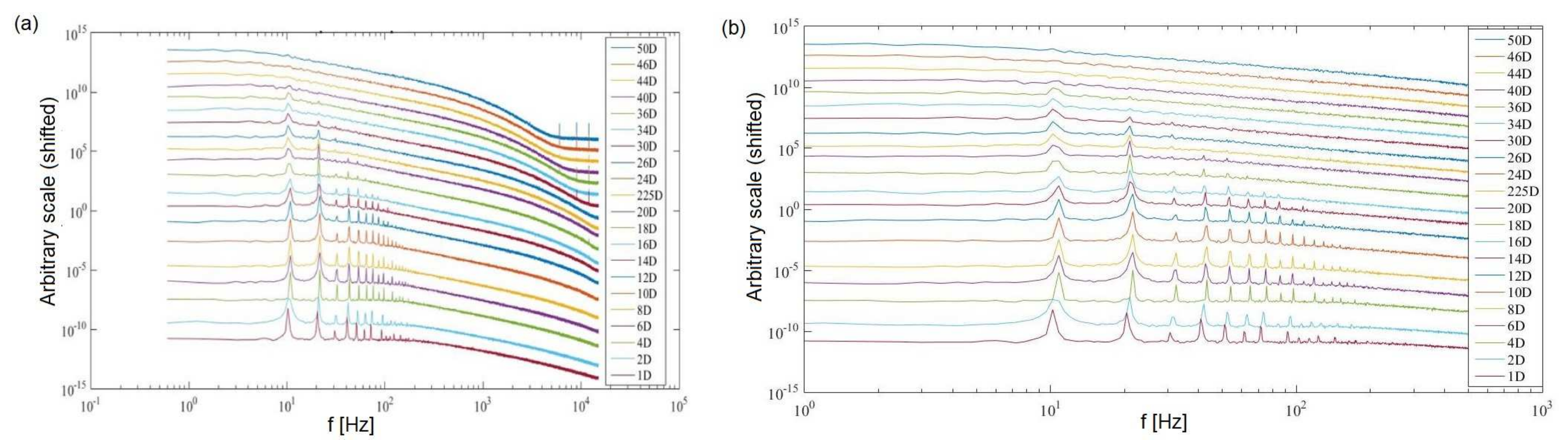}
\caption{Downstream spectrum development for $Re_D = 4.7 \cdot 10^4$ and $10D$ between the jet exit and the airfoil leading edge. (a) Full view and (b) zoom-in on the peak development.}
\label{fig:7}       % Give a unique label
\end{figure*}

An example of the hot-wire measurement results is displayed in Figure~\ref{fig:7}. This experiment was carried out at $Re_D = 4.7 \cdot 10^4$, keeping a distance of $10D$ between the jet exit and the leading edge of the airfoil. The spectra were measured at increasing axial distances between $1D$ and $50D$ from the airfoil trailing edge. These spectra are displayed in one graph with an offset of $10 - 15\, dB$ between each for the sake of comparison of the peaks.

Several features of this plot are noteworthy. It can be observed that the low order harmonics are created at an early stage, and several frequencies are observed already close to the trailing edge of the flap; note that the closest practical position of measurement was $1D$ from the airfoil trailing edge. Note that the third harmonic remains weaker than higher harmonics across the full downstream development. This phenomenon is explained in a companion paper~\cite{20} as a result of the finite region for the interaction of the participating Fourier components. The spectral window, caused by the finite interaction region, modifies the shape of the spectrum in comparison to a spectrum created in an infinite region. Furthermore, the generation of the third harmonic in particular requires the prior creation of the $2^{nd}$ harmonic in order to take place~\cite{20}.

Figure~\ref{fig:7} shows that the energy associated with the injected frequency of $10\, Hz$ depletes downstream, while the $20\, Hz$ second harmonic peak initially increases and subsequently decreases after reaching its maximum around approximately $8D$. A similar behavior can be seen for the higher harmonics.  The energy of the generated peaks eventually becomes absorbed in the developed turbulence further downstream. This behavior is similar to the one depicted in Figure~\ref{fig:6}: The transfer of energy appears to happen through a series of ``cascade events'' in agreement with the analysis in the companion papers~\cite{20,21}.

The striking stability and sharpness of the higher harmonics in Figure~\ref{fig:7} is noteworthy. Even at the late stages just before being absorbed, the positions of the peaks remain at their precise integer values of the airfoil excitation frequency and are apparently not smeared out by the surrounding, presumed Stochastic, turbulent velocity fluctuations. This behavior may partly arise due to the fact that the injected mode is characterized by a much larger energy than the ones relative to the general turbulence. Consequently, the interaction between harmonics wavenumbers remains more efficient, leading to the persistence of the initial structure far downstream in the jet (historically referred to as `permanence of large eddies'). This result lends strong support to the ideas of turbulence dependency upon initial conditions (in contrast to assuming universality), in line with our recent results~\cite{20,21}.

\begin{table}
% table caption is above the table
\caption{Reynolds number dependence of the time for absorption of the spectral peaks.}
\label{tab:1}       % Give a unique label
% For LaTeX tables use
\begin{tabular}{lll}
\hline\noalign{\smallskip}
$Re_D$ & Absorption time [s] \\
\noalign{\smallskip}\hline\noalign{\smallskip}
$2.2 \cdot 10^4$ & $0.1694$ \\
$3.0 \cdot 10^4$ & $0.1343$ \\
$3.6 \cdot 10^4$ & $0.1124$ \\
$4.2 \cdot 10^4$ & $0.0944$ \\
$4.7 \cdot 10^4$ & $0.0902$ \\
\noalign{\smallskip}\hline
\end{tabular}
\end{table}

The time for absorption of the peaks in Figure~\ref{fig:7} and corresponding parametric experimental investigations can be found from the integration of the downstream decaying velocity over the downstream distance. The results are summarized in Table~\ref{tab:1}. The absorption time of the peaks, which should be of significance to turbulence modelers, depends in the current case on the Reynolds number. In this regard, it has been noticed that the peaks are, independently of Reynolds number, consistently completely absorbed around approximately the same downstream spatial position, i.e. $\sim 44D$ after the trailing edge of the airfoil. This appears to be independent of the fact that the peaks from the non-linear interactions are more pronounced for higher Reynolds numbers.

\subsection{Vortex shedding in laminar jet core (HWA)}
\label{sec:4}

\begin{figure}
% Use the relevant command to insert your figure file.
% For example, with the graphicx package use
  \includegraphics[angle=270,width=\linewidth]{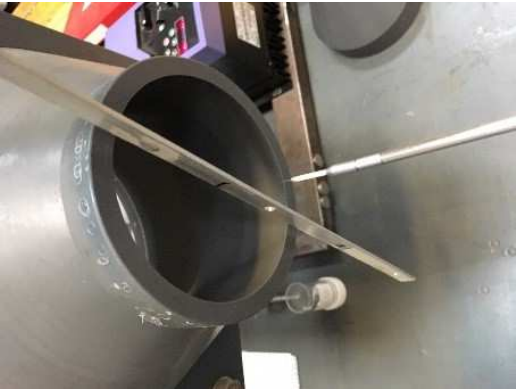}
% figure caption is below the figure
\caption{Jet with an extended laminar core, flat rod and hot-wire anemometer probe.}
\label{fig:8}       % Give a unique label
\end{figure}

Vortex shedding measurements were also carried out in the laminar core of a round turbulent jet with exit diameter $D=100\, mm$ and contraction ratio $2.4:1$. The jet could in this manner serve as an open wind tunnel with a nearly uniform exit velocity profile. The velocity can be determined from the pressure drop across the nozzle. A distinct vortex shedding frequency is injected into the laminar jet core by positioning a sharp-edged vertical rod at the nozzle outlet. The rod had a cross-section of $10\times 2 \, mm$. As Figure~\ref{fig:8} shows, the rod span extends across the entire nozzle diameter.

The objective was to isolate the workings of the nonlinear term by isolating the downstream development of a sharply defined Fourier mode injection using a shedding generator. This allowed for measurement of the very initial generation as well as the downstream development of the triadic interactions. This was not possible with the oscillating airfoil generated shedding, since measurements were only practical just behind the flap trailing edge where the energy distribution had already had ample time to develop across the airfoil chord. This setup, on the other hand, allows measurements from even the initial generation of the base frequency.

\begin{figure*}
% Use the relevant command to insert your figure file.
% For example, with the graphicx package use
  \includegraphics[width=\textwidth]{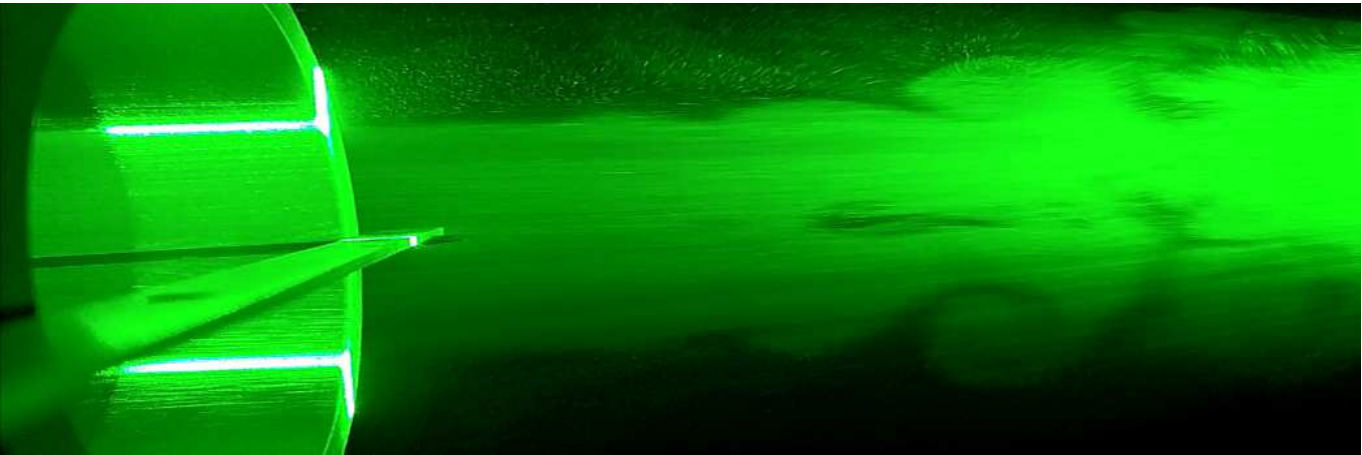}
% figure caption is below the figure
\caption{Vortex shedding flow visualization.}
\label{fig:9}       % Give a unique label
\end{figure*}

In Figure~\ref{fig:9} a flow visualization of the flow behind the vertical rod is shown. The laminar jet core around the sharp-edged rod is clearly visible in Figure~\ref{fig:9} along with the Kelvin-Helmholtz vortices developing farther downstream. Downstream of the rod, the flow vortex shedding formed by the sharp edge imposes a single fundamental frequency.

\begin{figure}
% Use the relevant command to insert your figure file.
% For example, with the graphicx package use
  \includegraphics[width=0.75\linewidth]{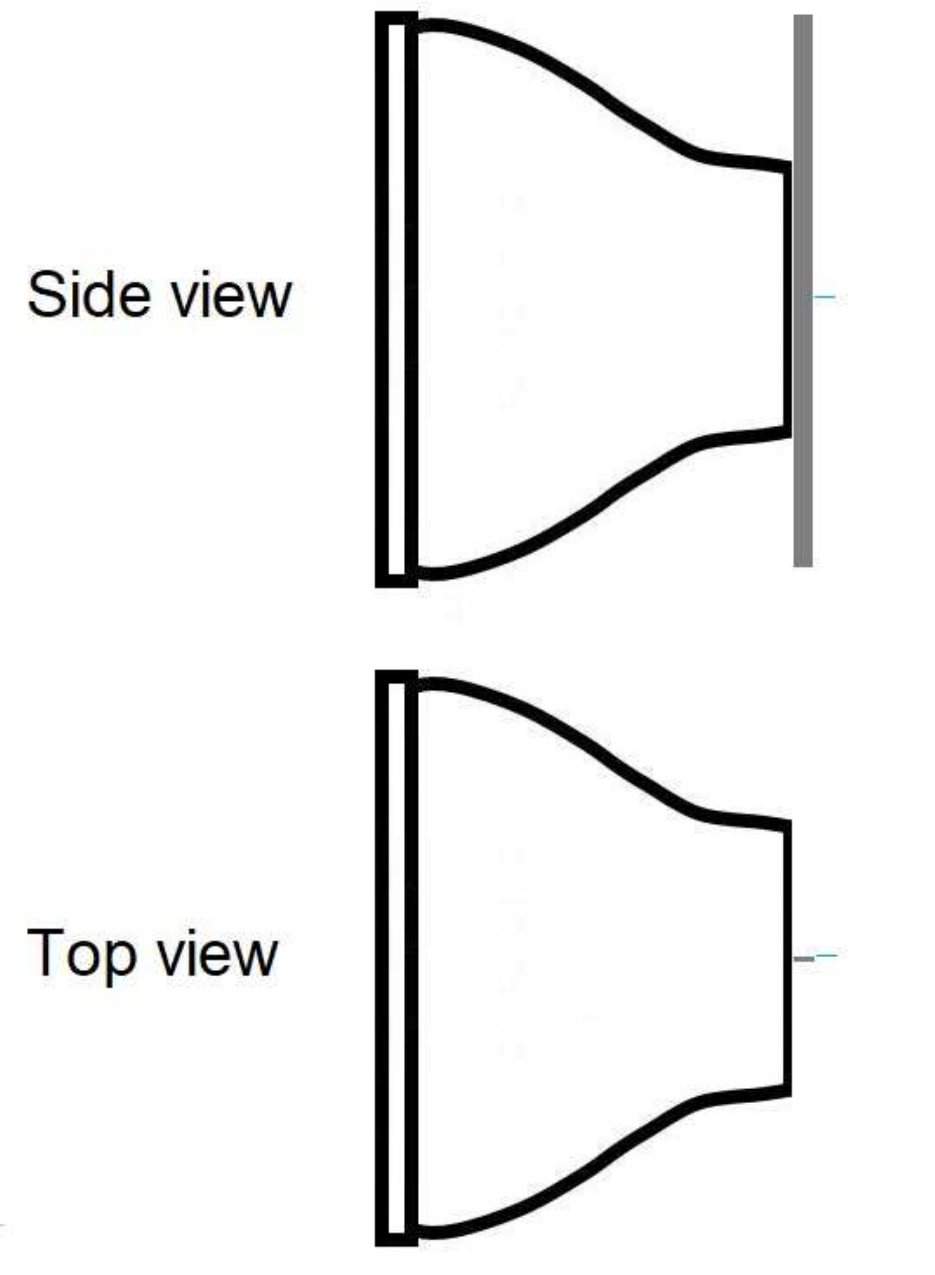}
% figure caption is below the figure
\caption{Schematic (not to scale) of the vortex shedding experiment in a laminar jet core, including the sharp-edged rod (in grey) to inject a sharp frequency and the region of the measurement points (indicated by a blue line).}
\label{fig:10}       % Give a unique label
\end{figure}

The same hot-wire anemometer system from Dantec Dynamics, as described previously, was employed. For the current experiment, the signal was captured using a PicoScope $5444B$ by Pico Technology. This scope allowed the change in both the waveform signal and the flow spectrum to be visualized in nearly real time, allowing therefore to quickly find identify the desired measurement points. The measurements were acquired with $0.5 \, mm$ intervals from $0.5\,mm$ to $10\, mm$ downstream from the vertical rod at $Re_D=1.06 \cdot 10^4$. Figure~\ref{fig:10} depicts a schematic of the linear region of the acquisition positions (in blue) from a top and a side view, respectively. The spanwise position was kept fixed to a distance of $1\, mm$ from the rod edge.

A total of $100$ time-series of $0.2\, s$ each were acquired in each measurement point with a sampling rate of $20\, kHz$, a hardware resolution of $12$ bits and sampling interval $50\, \mu s$. A spectrum was produced for each spatial point, cutting out the DC part of the signal and employing a Hamming window function, with a range of $50\, kHz$. A computer simulation applying a one-dimensional projection of Navier-Stokes equation onto the instantaneous flow direction is reported in~\cite{20}. The program assumes a time record of the velocity as input and computes the time development of the velocity trace employing multiple incremental passes through a fluid control volume exposed to the effect of the terms in the Navier-Stokes equation.

One result from this calculation is quoted and compared to one of the measurements. The measured time trace from the spatial point closest to the rod has been used as input to the computer program, so that the initial condition for the development of the velocity in the flow is identical to the initial velocity trace used in the program. The program uses a rectangular low-pass filter, which cuts out the DC part of the signal and some high frequency peaks due to external noise on the time traces. The spectra are then computed employing a Hamming window function, just like the PicoScope does. Note, the program uses only the terms in the Navier-Stokes equation without further approximations.

\begin{figure*}
% Use the relevant command to insert your figure file.
% For example, with the graphicx package use
  \includegraphics[width=\textwidth]{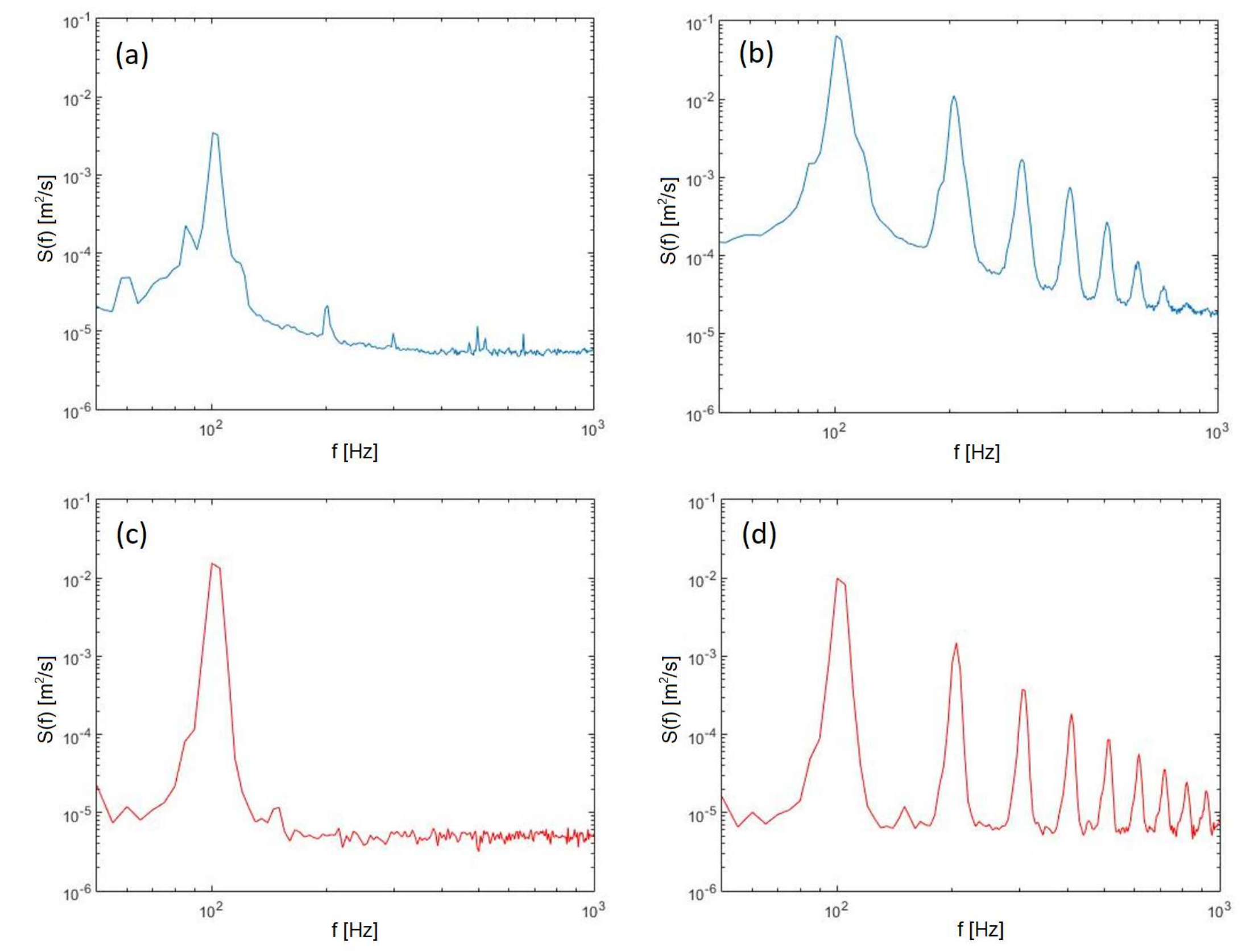}
% figure caption is below the figure
\caption{Comparison of velocity power spectra at $1\, mm$ (left) and $10\, mm$ (right) from the rod edge. Blue: Measurements, video available at https://doi.org/10.11583/DTU.12016827.v1. Red: Computer simulations, video available at https://doi.org/10.11583/DTU.12016869.v1.}
\label{fig:11}       % Give a unique label
\end{figure*}

Figure~\ref{fig:11} reports the downstream evolution of the experimental (blue) and the computational (red) downstream dynamic power spectra. The links in the figure captions provide moving images displaying the downstream evolution of the repeated workings of the nonlinear term on the initial modal injection. Note that the acquisition instrument introduces a filter, manifesting itself as a slow variation across frequency, which should be disregarded when interpreting the results.

The plots in Figure~\ref{fig:11}(a) and~\ref{fig:11}(b) display respectively the first spectrum measured close to the vortex shedding rod and the spectrum measured at a distance of $10\, mm$ downstream. Figures~\ref{fig:11}(c) and~\ref{fig:11}(d) display the corresponding results from the simulation. From these results, the delays in the cascade wavenumber interactions are particularly evident. Indeed, in contrast to Figure~\ref{fig:7}, where the low order harmonics are already present even in the most upstream measurement, in Figure~\ref{fig:11} only the initial peak is present at the initial measurement point.

The time scale for vortex shedding to develop and reach an `asymptotic' configuration (like the one shown in Figure~\ref{fig:11}b) has been computed for the considered experiment and was found to be $6\, ms$. The time was estimated from the integration of the downstream velocity over the downstream distance, as for the time for absorption of the peaks of Table~\ref{tab:1}.

\section{Conclusions}
\label{sec:5}

The energy transfer between different scales of velocity fluctuations is a key process in the development of turbulence, and knowledge of the efficiency and time scales for these energy exchanges is crucial for the understanding of turbulence theory and for further development of engineering models. The exchange between harmonics of the originally injected frequency is well in line with the fact that frequency doubling and sum- and difference-frequencies are to be expected as a result of the actions of the nonlinear term of the Navier-Stokes equation~\cite{21}.

Three experiments were performed where a single Fourier mode was injected into a well-defined turbulent flow and the development of the velocity power spectrum was traced along the downstream direction. The downstream convection has been considered as a successive exposure of the initial velocity trace to the Navier-Stokes equation in a small fluid control volume. This point of view allowed us to compare the development of the measured spectra to the spectra computed by a one-dimensional computer simulation with the measured initial time trace as input and to compute the power spectra as this time trace was transformed by incremental repeated exposures to the Navier-Stokes equation.

Although the particular form of the measured and computed power spectra depends on the initial position of the measurement point and the way the Fourier mode was injected, a number of common properties were revealed:
\begin{itemize}
  \item Higher order frequency components were formed successively as a fluid element passing near the point of injection evolved downstream in time, as expected from the cascade process described  in~\cite{20}. 
  \item The higher frequencies were exact multiples of the injected frequency and they retained their well-defined sharp spectral frequency -- even when submerged into high-intensity turbulence.
  \item Far downstream the spectrum tends to an asymptotic form whose energy is slowly being absorbed into the surrounding turbulence. This agrees with results in~\cite{21}.
  \item The absorption distance does not change significantly with Reynolds number in the investigated range $Re = 22.000 - 47.000$. The absorption time, consequently, varies with Reynolds number (jet exit velocity).
  \item The time development of the generation of the higher harmonics is evident in all the measurements, but is most clearly seen in the vortex shedding experiment and the agreement with the numerical simulations in~\cite{21}.
  \item Influence of the initial spectral components on the far downstream spectrum is clearly evident, in particular in the oscillating airfoil measurements in the highly turbulent jet.
\end{itemize}

All in all, the measurements and computer simulations illustrate the dynamic nature of the triadic interactions between injected energetic Fourier components through the downstream development of the velocity and power spectrum, as predicted from~\cite{20,21}.

The authors hope that these results can be used by turbulence modelers to improve their methods, in particular the results on the generation and absorption of the harmonics of the initial injected frequency.

\begin{acknowledgements}
Benny Edelsten is acknowledged for his assistance with the hot-wire experiments.
\end{acknowledgements}

 \section*{Conflict of interest}

 The authors declare that they have no conflict of interest.

\end{document}